# Foreign Portfolio Investment and Economy: The Network Perspective

Muhammad Mohsin Hakeem*[1]
Doctoral Fellow, Graduate School of Economics and Management, Tohoku University,

Ken-ichi Suzuki
Associate Professor, Graduate School of Economics and Management, Tohoku University, Sendai, Japan

**Abstract**
The European Union and Eurozone present an inquisitive case of strongly interconnected network with high degree of dependence among nodes. This research focused on investment network of European Union and its major trading partners for specific time period (2001-14). The changing investment patterns within Eurozone suggest strong financial and trade links with central and large economies. This study is about the association between portfolio investment and economic indicators with respect to financial networks. The analysis used the strongly connected investment network of Eurozone and its large trading partners. A strong correlation between, increasing or decreasing investment patterns with economic indicators of particular economy was found. Interestingly correlation patterns for network members other than Eurozone states were not as strong and depicted mild behavior. This as well, explains the significance of interconnectedness level among nodes of one network with varying centrality measures. Investment network visualization techniques helped to validate the results based on network's statistical measures.

**Keywords:** European Union, Eurozone, Investment Network, Economic Indicators, Centrality Measures, Network Visualization

**INTRODUCTION**

Portfolio investment is one of the major indicators of investor friendly and good performing equity markets of a single country. The rate of return is for sure the most prominent factor behind investment decision but ease of access, financial stability and lower level of taxation do play evident role in final decision of investment managers regarding liquidity flows. Being a part of an investment network, either week or strong, can open up new possibilities to attract foreign investors by making markets more visible and investor friendly. The networks such as European Union (EU) or Eurozone are supposed to influence the investment flows for any particular nodes within, due to strong connectivity patterns and possibility of small clusters. This study explores the details of individualistic or local characteristics of every node within investment network. Mainly the connectivity patterns, closeness within network or possibility of large nodes in neighborhood. All these characteristics can influence the portfolio investment flows for any particular country. The objective of this study is to analyze the connectivity patterns with network and major economic indicators of any country to find the possible connection or correlation in-between. In other words, if economic stability or deterioration with respect to certain indicators can be associated with strong or weak connectivity patterns with network as a resultant economic state of affairs.

Association or linkage does not have implications for causality. The strong linkage with network and higher or improved economic indicators may or may not represent the underlying causes but would reflect on different possibilities. To understand these phenomena we used investment network of Europe Union with focus on individualistic characteristics of nodes. Based on criteria concerning closeness and connectivity nodes were divided into different Tiers. At least one node from every Tier was selected to build correlation matrix based on networks measures and relevant economic indicators to understand the relationship between network position, economic stability and attractiveness for investors. The network is based on coordinated portfolio investment survey (CPIS) database compiled and published by International Monetary Fund (IMF) regularly.

---

[1] "hakeem514@tohoku.ac.jp"

**RELATION TO THE LITERATURE**

The discussion on the relationship between portfolio investment and economy is not new rather a continuous process of development, enhanced understanding and evaluation of different perspectives. Portfolio investment and associated concepts are such as, its determinants, investor protection, efficiency of capital markets, flow determinants and patterns, exchange rate movements and information mobility are few to mention. The economic relations are also part of widely available literature. Such as Rogoff (1999) focused on considerable change from debt to equity financing within economy and equity investment flows increases accordingly. Bekaert & Harvey (1998) confirm the direct impact of private equity investment on macroeconomic performance of emerging market. They also confirmed the impact of portfolio investment on economic growth and stability of emerging economies in Bekaert & Harvey (2000). Their later paper Bekaert & Harvey (2003) confirms the impact of increased liquidity and better access to cheaper financing on economic activity of host country.

Researchers such as Levine & Zervos (1996) discussed the investment's impacts on liquidity and implications for better and broader market. The issues related to improvement of foreign portfolio investment for any country and its contribution towards more efficient stock market and elimination of financial constraints for domestic corporations are discussed in detail by Laeven (2003) and Knill (2004). Besides the positive impacts of portfolio investment there are studies focusing on short or long term adverse effects besides betterment of capital markets and capital access. The multiplier effect for the growth of capital markets improves the liquidity situation for all investors; the capital flows are the depiction of enhanced economic growth and activity and add value towards wealth creation and distribution. Efficient capital allocation is the ultimate point which can help the host economy to grow multidimensional and dynamically. Rajan and Zingales (1998); Wurgler (2000) and Love (2003) contributed for better explanation of these issues. There are studies focusing on economic development of different countries due to foreign portfolio inflows such as Agarwal (1997) focused on Korea, Indonesia, India and Thailand while Duasa & Kassim (2009) on Malaysia. Both studies concluded on positive note in terms of relationship between portfolio investment and economy.

Beside wide spectrum of literate on portfolio investment flows, resultant efficiency of markets and economic impact there is scarcity of network perspective especially liquidity flows and resultant impacts. There are studies related to network analysis of capital markets such as network analysis of Chinese stock market by Huang, Zhuang & Yao (2009) and few others. The transformation process of investment network is discussed in Hakeem & Suzuki (2016a and 2016b). We extended our analytical approach to evaluate the characteristics of individual nodes to establish the relationship between portfolio investment flows and economic indicators.

**THEORETICAL BACKGROUND**

The distinct characteristics of nodes are explored by using multiple centrality and relevant measures at micro level. Closeness centrality, clustering coefficient, in-degree and out degree is examined for every single node to categories them accordingly. By using the criteria based on these centrality and other measures we divide the existing nodes into three classes or Tiers. Steps of analytical process are as follows.
1. Analysis of individual characteristics of nodes by using centrality and relevant measures.

2. Classification criteria based on resultant measures
3. Classification of 26 nodes into 3 different Tiers based on their individual positions within network
4. Selection of at least one nodes from each Tier for correlation analysis
5. Selection of macroeconomic indicators for designated countries
6. Correlation matrices based on network analytics and macroeconomic indicators for designated countries
7. Identification of correlation patterns and differences according to nodes and Tier classifications

**Centrality Measures**

Following centrality measures are used for investment network to understand individualistic characteristics of nodes within. We would introduce few measures briefly; detailed description of centrality measure and implications for network analysis is available in Hakeem and Suzuki (2015).

*Degree Centrality*

The simplest and earliest centrality measure in a network is the degree of a node, the number of edges connected to it. In directed networks nodes have both an in-degree and an out-degree, and both may be effective if used in the appropriate circumstances. Although degree centrality is a simple centrality measure, it can be very insightful. In a financial network, for instance, the financial institution or a node connected to all other nodes can have much more influence on other nodes as well as the resilience of whole network. The standardized degree centrality of a node is its degree divided by the maximum possible degree.

$$c_i^d = \frac{d}{n-1} \quad (1)$$

Aggregate degree centrality for the whole network is

$$C^d = \frac{\sum_{i=1}^{n} |c_i^d - c_i^d *|}{(n-2)(n-1)} \quad (2)$$

Where *degree centrality* "$C^d$" is calculated by using the maximum value, while *n* represents the number of nodes within that particular network. The higher the number of nodes the higher degree centrality it can have. The degree centralization of any regular node is 0, while star has degree centralization of 1.

For a node the number of edges ends with it as known as in-degree and number of edges originating from it is known as out-degree. The node with no in-degree but all out degrees is known as "source" and the one with all in-degrees but no out-degree is called "sink". A balanced directed graph has equal number of in and out degrees.

*Closeness Centrality*

This centrality measure is totally different, as it measures the mean distance from one node to other nodes. It is the concept of geodesic path, - the shortest path between two nodes-. Closeness centrality has small values for nodes that are separated from others by only a short geodesic distance on average. Such nodes might have better access to information at other nodes or more direct influence on other nodes. In a financial network, for example, a financial institution with lower mean distance to others might have better access to liquidity and important financial information. Closeness centrality is a very natural measure of centrality and is often used in different types of network studies. Closeness is based on the length of the average shortest path between a vertex and all vertices in the graph

$$C_i^c = \frac{n-1}{\sum_{j \neq i}^{n-1} \delta_{ij}} \tag{3}$$

Where $\delta_{ij}$ represent the geodesic path between i and j. Aggregate centrality for the whole network can be defined as follows.

$$C^c = \frac{\sum_{i=1}^{n} |C_i^c - C_i^c *|}{(n-2)(n-1)(2n-3)} \tag{4}$$

If $C_i^c *$ is the maximum closeness centrality a node attained, then the aggregate closeness centrality is the variation in closeness centrality of all nodes divided by maximum possible closeness centrality for particular network.

Whereas Normalized Closeness Centrality is,

$$C_i^{c\prime} = C_i^c / (n-1) \tag{5}$$

Where, $\delta_{ij}$ is the distance between node *i* and *j*, while *n* refers to the number of nodes within network.

*Clustering Coefficient*

The clustering coefficient is the degree by which nodes tends to make groups or clusters together. The clustering of nodes having similar connectivity patterns or others characteristics is evident in network analysis. There are two ways to measure the clustering of nodes in particular networks.

1. Global Clustering Coefficient
2. Local Clustering Coefficient

This first type "Global Clustering Coefficient" is based on trio of nodes. The trio is combination of three nodes connected to each other. The clustering coefficient measures the density of triangles in the network.

$$C^{Cl} = \frac{1}{n} \frac{[(k^2) - (k)]^2}{k^3} \tag{6}$$

In a random network of connections between nodes and edges, $k^2$ and $k$ has fixed or finite values the quantity becomes small as $n \to \infty$, so the clustering coefficient can be small as size of network grows. But in reality it can be very different depending on network type and size. The aggregate clustering coefficient can be calculated by taking the mean of local clustering coefficient of each node.

$$C^{Cl} = \frac{1}{n} \sum_{i=1}^{n} C_i^{cl} \tag{7}$$

Whereas the local clustering coefficient of a node can be defined as follows,

$$C_i^{cl} = \frac{e_{jk}}{k_i(k_i - 1)} \tag{8}$$

While $e_{jk}$ is the path from i to j, and $k_i$ are the number of neighbors of a node. We can also represent it in the following way.

$$C_i^{cl} = \frac{n_i}{k_i(k_i - 1)} = \frac{\sum_{jk} e_{ij} e_{jk} e_{ki}}{k_i(k_i - 1)} \tag{9}$$

**The Correlation**

Any statistical relationship among two random variables can be termed dependence or linkage in network context. Correlation involves dependence or linkage among two variables. Though, in statistical context it is the level of linear relationship among two variables. A simple example of correlation can be the relationship between supply and price of crude oil in international market. As supply increases the price goes down accordingly.

We used "Pearson's product moment correlation coefficient" to explain the linkage between network indices and economic indicators. By consideration the basic difference between correlation and causation we developed matrices to analyze the linkage for different countries during varying time periods.

For series of n measurements of X and Y, known as $x_i$ and $y_i$ for $i = 1, 2, \ldots \ldots n$, the sample correlation coefficient can be used to estimate correlation $r_{xy}$ between both variables. It can be written as,

$$r_{xy} = \frac{\sum_{i=1}^{n}(x_i - \bar{x})(y_i - \bar{y})}{n s_x s_y} = \frac{\sum_{i=1}^{n}(x_i - \bar{x})(y_i - \bar{y})}{\sqrt{\sum_{i=1}^{n}(x_i - \bar{x})^2 \sum_{i=1}^{n}(y_i - \bar{y})^2}} \quad (10)$$

Where $\bar{x}$ and $\bar{y}$ are the sample means and $s_x$ and $s_y$ are the sample standard deviations of X and Y. We can also write it as follows,

$$r_{xy} = \frac{\sum x_i y_i - n \bar{x} \bar{y}}{n s_x s_y} = \frac{n \sum x_i y_i - \sum x_i \sum y_i}{\sqrt{n \sum x_i^2 - (\sum x_i)^2} \sqrt{n \sum y_i^2 - (\sum y_i)^2}} \quad (11)$$

The correlation coefficient can be +1 as to represent perfectly positive correlation relationship or -1 to show perfectly negative correlation among two variables. Range of resultant matrix is $-1 \leq r \leq +1$, which can explain the strength, level and type of relationship.

## EXPLORING THE NETWORK
### The Investment Network

In our analysis of investment network we used the data of Coordinated Portfolio Investment Survey (CPIS) compiled and published regularly by IMF. The same data set is widely used in literature for network or non-network analysis related to global portfolio investment patterns. The CPIS data is aggregate amount received by single country or invested in one country by foreign individuals, corporations and investment agencies or other vehicles in equity markets. We used data from 2001 to 2014, total number of 14 years. There are 26 countries or nodes within this network. Out of these 26 countries 24 are European Union (EU) members while 2, US and Japan, are major partners of EU in investment and trade. There are 28 EU member states; our sample includes all major and prominent nodes according to economic output and capital market statistics. The countries excluded due to data constraints are Latvia, Lithuania and Slovakia as Eurozone member states and Croatia as EU member.

The timespan selected is interesting as we have seen huge ups and downs within this decade and we can easily call it a decade of change. There was Global Financial Crisis (GFC) impacting housing, equity and debt markets directly, initiating around late 2007. Europe also faced a daunting rather tough debt crisis starting after GFC. The debt crisis compressed weak European economies and had severe impacts on bilateral relations within European Union. The resultant austerity measures impacted millions of households in effected countries by increasing direct and indirect taxes, reducing employment opportunities and hampering growth and development.

### Selection Criteria and Nodes

By considering the centrality measures and observing the flow patterns, we categorized all nodes into three different Tiers.
- Tier 1 - Strong level of connectivity
- Tier 2 - Intermediate level of connectivity
- Tier 3 - Low level of connectivity

**Table 1: Average Closeness Centrality and Classification, Source (Authors Calculation)**
**CC: Closeness Centrality, n= number of nodes**

| Classification Based on Average Closeness Centrality | | |
|---|---|---|
| Tier 1 | Tier 2 | Tier 3 |
| $CC \leq 1.05$ | $CC \leq 1.20$ | $CC \geq 1.21$ |
| $n = 13$ | $n = 9$ | $n = 4$ |
| Austria Luxembourg United States Germany Italy France Netherlands Belgium Ireland Denmark Sweden Cyprus Japan | United Kingdom Estonia Czech Republic Greece Slovak Republic Hungary Finland Spain Portugal | Poland Bulgaria Malta Romania |

Table 1 explains the classification of all nodes with respect to average closeness centrality. There are 13 nodes in first Tier, which represent the strongly connected nodes of investment network. The inclusion of these nodes within first Tier is confirmed by in and out-degree measures as well. There isn't any surprise inclusion within this Tier as connectivity and flow of all relevant countries is high enough. Tier 2 represents mid-level connectivity of included nodes with rest of the network. There is a surprise inclusion within this group, the United Kingdom. London being the capital of UK is hub of international bond market. LIBOR is used worldwide for settlements of debt and relevant contracts. Though UK is just above the criteria and with small relaxation it can join Tier 1 countries. Our early data indicates the lower level of in and out degree measures for UK for certain time period though. As being an attractive investment destination it might not be able to invest into other foreign markets. Rest of Tier 2 countries follows their intermediate connectivity levels within network. Tier 3 includes the least central nodes with less connectivity and flow with other partners. These countries tend to have higher clustering coefficient as they are not fully connected with the whole network. We believe this group is balanced and accurate according to degree, clustering and centrality measures.

We would select four countries for further analysis to establish our theory about connectivity and economy. The following countries are selected from all tiers,
1. Germany (Tier 1)
2. France (Tier 1)
3. Greece (Tier 2)
4. Romania (Tier 3)

The countries are selected according to the connectivity levels as are represented in table 1.

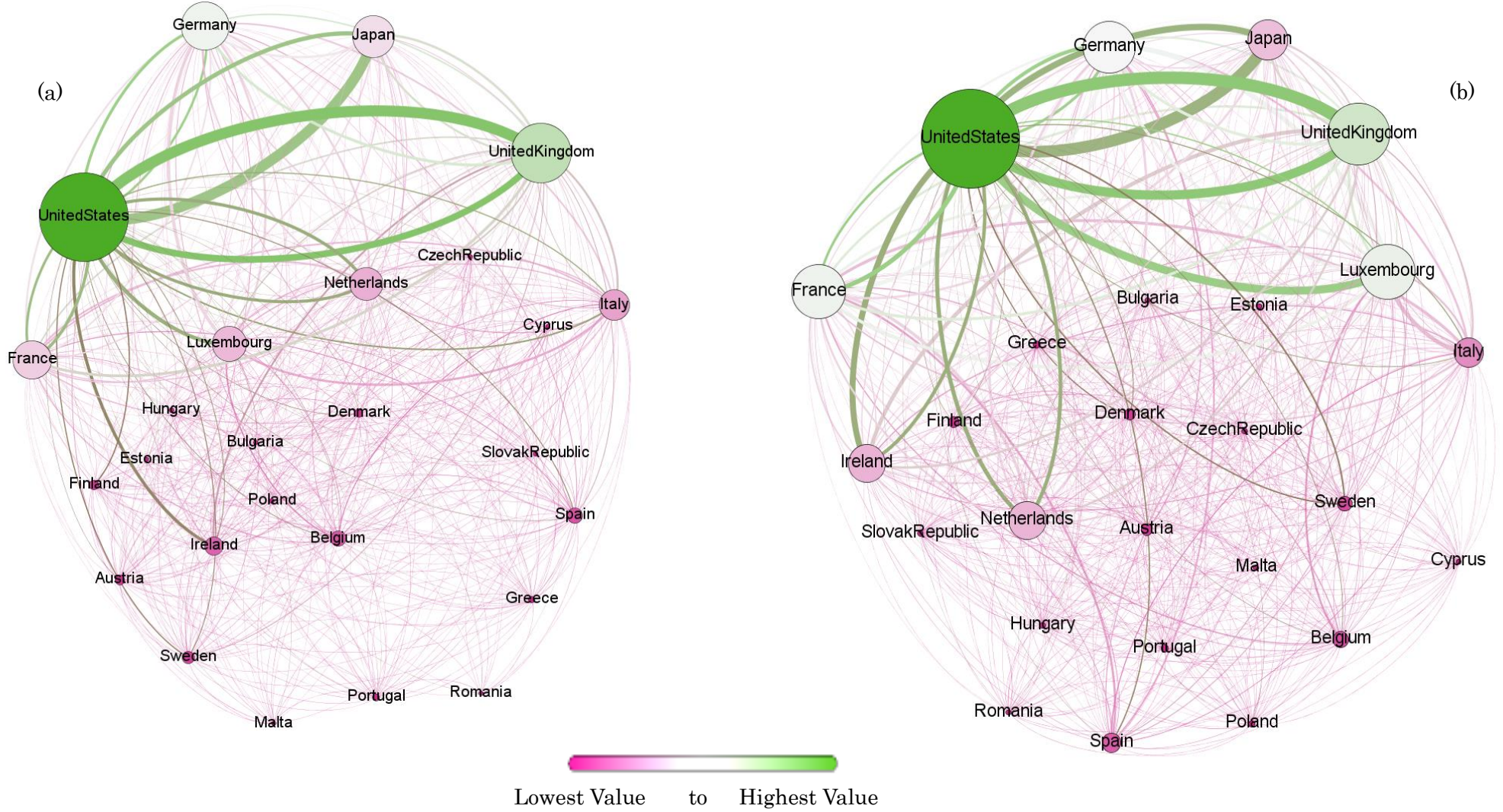

*Figure 1: Investment Network for the year 2001(a) and 2014(b)
(Source: Coordinated Portfolio Investment Survey, International Monetary Fund 2015)*

Figure 2: Correlation Matrices of France (a), Germany (b), Greece (c) and Romania (d)
N1 to N11 represent network indices while E1 to E8 represent economic indicators
Color Patterns: Red represents negative correlation; Green represents positive correlation and yellow represent data point below threshold value

(a)

| | E1 | E2 | E3 | E4 | E5 | E6 | E7 | E8 |
|---|---|---|---|---|---|---|---|---|
| N1 | 0.593 | 0.539 | 0.594 | (0.533) | 0.521 | 0.373 | 0.457 | (0.566) |
| N2 | 0.884 | 0.830 | 0.886 | (0.794) | 0.783 | 0.418 | 0.683 | (0.814) |
| N3 | 0.907 | 0.844 | 0.909 | (0.815) | 0.801 | 0.468 | 0.700 | (0.844) |
| N4 | 0.936 | 0.943 | 0.926 | (0.930) | 0.913 | 0.624 | 0.849 | (0.873) |
| N5 | 0.895 | 0.952 | 0.875 | (0.959) | 0.946 | 0.746 | 0.912 | (0.861) |
| N6 | 0.946 | 0.904 | 0.944 | (0.873) | 0.853 | 0.488 | 0.762 | (0.857) |
| N7 | (0.870) | (0.843) | (0.866) | 0.813 | (0.798) | (0.501) | (0.701) | 0.847 |
| N8 | (0.884) | (0.830) | (0.886) | 0.794 | (0.783) | (0.418) | (0.683) | 0.814 |
| N9 | (0.785) | (0.766) | (0.784) | 0.739 | (0.737) | (0.376) | (0.672) | 0.713 |
| N10 | 0.940 | 0.910 | 0.937 | (0.891) | 0.884 | 0.549 | 0.806 | (0.874) |
| N11 | 0.936 | 0.943 | 0.926 | (0.930) | 0.913 | 0.624 | 0.849 | (0.873) |

(b)

| | E1 | E2 | E3 | E4 | E5 | E6 | E7 | E8 |
|---|---|---|---|---|---|---|---|---|
| N1 | 0.836 | 0.665 | 0.827 | (0.815) | 0.798 | 0.694 | 0.660 | 0.872 |
| N2 | 0.763 | 0.571 | 0.751 | (0.722) | 0.571 | 0.573 | 0.565 | 0.712 |
| N3 | 0.850 | 0.658 | 0.839 | (0.818) | 0.734 | 0.676 | 0.653 | 0.845 |
| N4 | 0.929 | 0.854 | 0.927 | (0.938) | 0.853 | 0.883 | 0.832 | 0.958 |
| N5 | 0.902 | 0.768 | 0.895 | (0.887) | 0.833 | 0.801 | 0.747 | 0.930 |
| N6 | 0.932 | 0.898 | 0.932 | (0.957) | 0.852 | 0.923 | 0.874 | 0.961 |
| N7 | (0.763) | (0.571) | (0.751) | 0.722 | (0.571) | (0.573) | (0.565) | (0.712) |
| N8 | (0.763) | (0.571) | (0.751) | 0.722 | (0.571) | (0.573) | (0.565) | (0.712) |
| N9 | (0.824) | (0.790) | (0.831) | 0.847 | (0.797) | (0.856) | (0.826) | (0.839) |
| N10 | 0.932 | 0.825 | 0.930 | (0.937) | 0.838 | 0.856 | 0.833 | 0.947 |
| N11 | 0.929 | 0.854 | 0.927 | (0.938) | 0.853 | 0.883 | 0.832 | 0.958 |

(c)

| | E1 | E2 | E3 | E4 | E5 | E6 | E7 | E8 |
|---|---|---|---|---|---|---|---|---|
| N1 | 0.840 | 0.736 | 0.842 | (0.289) | 0.640 | (0.105) | 0.628 | (0.512) |
| N2 | (0.242) | (0.121) | (0.234) | (0.145) | (0.003) | 0.176 | (0.141) | 0.590 |
| N3 | 0.741 | 0.677 | 0.745 | (0.325) | 0.620 | (0.048) | 0.566 | (0.317) |
| N4 | 0.852 | 0.435 | 0.852 | 0.151 | 0.254 | 0.131 | 0.246 | (0.727) |
| N5 | 0.679 | 0.083 | 0.677 | 0.556 | (0.142) | 0.193 | (0.105) | (0.824) |
| N6 | 0.536 | 0.802 | 0.540 | (0.778) | 0.850 | (0.096) | 0.757 | 0.033 |
| N7 | (0.153) | (0.310) | (0.154) | 0.277 | (0.323) | 0.284 | (0.425) | 0.141 |
| N8 | 0.242 | 0.121 | 0.234 | 0.145 | 0.003 | (0.176) | 0.141 | (0.590) |
| N9 | (0.031) | (0.539) | (0.034) | 0.819 | (0.681) | 0.003 | (0.664) | (0.374) |
| N10 | 0.658 | 0.884 | 0.663 | (0.762) | 0.917 | (0.069) | 0.859 | (0.099) |
| N11 | 0.852 | 0.435 | 0.852 | 0.151 | 0.254 | 0.131 | 0.246 | (0.727) |

(d)

| | E1 | E2 | E3 | E4 | E5 | E6 | E7 | E8 |
|---|---|---|---|---|---|---|---|---|
| N1 | 0.615 | 0.616 | 0.606 | 0.037 | 0.663 | 0.598 | 0.411 | (0.340) |
| N2 | 0.914 | 0.956 | 0.923 | (0.231) | 0.942 | 0.804 | 0.799 | (0.186) |
| N3 | 0.919 | 0.951 | 0.922 | (0.159) | 0.958 | 0.830 | 0.755 | (0.269) |
| N4 | 0.736 | 0.853 | 0.755 | (0.534) | 0.876 | 0.659 | 0.909 | 0.120 |
| N5 | 0.697 | 0.821 | 0.716 | (0.557) | 0.846 | 0.617 | 0.901 | 0.164 |
| N6 | 0.894 | 0.954 | 0.903 | (0.338) | 0.959 | 0.837 | 0.855 | (0.162) |
| N7 | 0.456 | 0.481 | 0.451 | 0.173 | 0.521 | 0.309 | 0.275 | (0.263) |
| N8 | (0.914) | (0.956) | (0.923) | 0.231 | (0.942) | (0.804) | (0.799) | 0.186 |
| N9 | 0.741 | 0.622 | 0.730 | 0.125 | 0.611 | 0.776 | 0.375 | (0.507) |
| N10 | (0.210) | (0.234) | (0.207) | (0.188) | (0.256) | 0.046 | (0.073) | 0.073 |
| N11 | 0.736 | 0.853 | 0.755 | (0.534) | 0.876 | 0.659 | 0.909 | 0.120 |

# CORRELATION BETWEEN NETWORK AND ECONOMIC INDICATORS

**Economic Indicators**

Economic Indicators used for analysis are obtained from International Monetary Fund (IMF). The database of International Financial Statistics (IFS) is used to obtain the relevant measures. The IFS is one of the fund's main databases and is available since 1948. We used the similar timespan as we have for our Network data, from 2001-2014. Total numbers of economic indicators obtained and used in analysis were in double digit, a rough estimate stands around 30. At the final stage 8 economic indicators were selected. To elaborate our idea of correlation between investment network and economy we used the following economic indicators.

Table 2: Economic Indicators used for Analysis (Source: IFS 2015, IMF)

| S. No. | Economic Indicators | Code Assigned |
|---|---|---|
| 1 | Gross domestic product, current prices | E1 |
| 2 | Gross domestic product, deflator | E2 |
| 3 | Gross domestic product per capita, current prices | E3 |
| 4 | Gross domestic product based on purchasing-power-parity (PPP) share of world total | E4 |
| 5 | Inflation, average consumer prices | E5 |
| 6 | General government revenue | E6 |
| 7 | General government gross debt | E7 |
| 8 | Current account balance | E8 |

**Network Indicators**

The network indicators used for analysis are obtained from CPIS (Coordinated Portfolio Investment Network) investment network. The CPIS database is compiled and published by IMF regularly. The network indicators or indices are outcome of our calculations unlike the economic indicators which are available through database. The methodology and calculation mechanism is briefly explained here, for detailed theoretical background please refer to Hakeem and Suzuki (2015).

Table 3: Network Indicators used for Analysis (Source: CPIS 2015, IMF)

| S. No. | Network Indicator | Code Assigned |
|---|---|---|
| 1 | In-Degree | N1 |
| 2 | Out-Degree | N2 |
| 3 | Degree | N3 |
| 4 | Weighted Degree | N4 |
| 5 | Weighted In-Degree | N5 |
| 6 | Weighted Out-Degree | N6 |
| 7 | Eccentricity | N7 |
| 8 | Closeness Centrality | N8 |
| 9 | Betweenness Centrality | N9 |
| 10 | Clustering Coefficient | N10 |
| 11 | Strength | N11 |

**Correlation Matrices of Selected Countries**

The correlation matrices are presented in figure 2 (a) to (d) for France, Germany, Greece and Romania respectively. These matrices are based on the 14 years data of economic indicators and network indices. These matrices give us insights regarding relationship or linkages between portfolio investment and economic conditions of certain countries. The results can be generalized for other countries in the similar "Tier".

*Tier 1 Countries*

Tier 1 is representative of countries having strong connectivity and flow linkages with network. The group is composed of 13 countries of which 11 belong to EU besides 2 countries, Japan and US, representing others. The correlation matrix for the France and Germany is not identical like their overlapped network indices. But similarities are in much larger percentage compared to differences. First we would look at figure 2a, which represent France's matrix. There is indeed a correlation pattern and relationship between networks indices and macroeconomic indicators.

There is a strong correlation between Gross Domestic Product (GDP), Purchasing Power Parity (PPP) with In-Out degree and weighted degree measures. It means that higher the connectivity level higher the impact on economic growth patterns. There is a strong negative correlation between network's centrality measures such as closeness, betweeness and eccentricity with GDP and PPP. This negative correlation depicts the positive impact due to technical reason; the centrality measures move to reverse side or statistically decreases if level of centrality improves. It means the more central nodes would have lower values compared to less central nodes. The correlation matrix takes this on the opposite side. As if GDP is increasing and statistics of closeness are decreasing it makes a perfect sense of negative correlation. The implications of this strong negative correlation are positive. So the relationship between closeness, betweeness and eccentricity is strong and understandable. There are two other strong relations between network indices, inflation and current account balance. It is interesting to know that the strongly connected nodes have less inflation and better trade relations with trading partners. Cumulatively for France there are strong relations between its investment network and economic indicators. The relation is not causal rather exist; there can be more underlying reason besides the one under consideration in this study.

For Germany the relationship is of similar nature, the increase in centrality measures has strong connections with economic growth, inflation, current account balance and government debt. The current account balance of Germany is improving with the passage of time so is its centrality. That makes it positive correlation compared to France which has negative correlation with this particular variable. Tier 1 countries have important position within network; with strong centrality indices we can find correlation patterns with their economic indicators. These correlations are not causal rather shows the existence of relationship. The general conclusion should include this trend. The more central a country is, there more relationships or correlations it can have. It can also have implications for countries with less connectivity to improve their network position and capture more economic benefits.

*Tier 2 Countries*

Tier 2 countries are modestly linked with investment network. Greece is selected as representative of this group. The country has modest linkages with investment network. It showed improvements regarding connectivity patterns initially, had good position for a while and again feeling the heat of Eurozone debt crisis.

The correlation matrix representing the possible ties between Greece's economy and network position is shown in figure 2c. The correlation patterns are there, but if we compare it with Tier 1 countries then the level of correlations is much lower and scattered. We might not be able to conclude of having any strong relationship between economic indicators and network indices. Being a standalone country and group representative of Tier 2 we can find certain patterns for certain period of time at least. The connection of In-out degree and weighted degree with GDP and PPP is one of the prominent one. There is also an indication of link between inflation and general government debt with clustering coefficient. Tier 2 countries have higher possibility of joining any cluster compared to Tier 1. So the explanation regarding increase in clustering coefficient due to Eurozone debt crisis might have exposed Greece with better inflation management and government debts.

The conclusion on Greece can be generalized for Tier 2 countries due to economic similarities and prevailing circumstances. These nodes do not have strong connectivity pattern, so experience variations in their position within network. The links or correlation between network indices of Tier 2 countries and economic indicators are not high enough, due to limited connectivity and exposure.

*Tier 3 Countries*

Tier 3 countries are weakly linked with investment network and do not possess a strong position within it. Romania is the representative of this small group which may feel or is alienated compared to Tier 1 and Tier 2 countries. Romania is an interesting case of connectivity. It has extremely lower level of connectivity initially, and improved at later stage. Recently it reached the same degree connectivity level as Greece. So we would be able to analyze the changes in connectivity and consequences on relationship.

Figure 2d represents the Romania's correlation matrix. It seems to have modest level of correlations between economic indicators and network indices. If compared with Tier 1 then the patterns are not that significant. But for Tier 2 it's not that weak either. It seems Romania may have better relation between its economic indicators and networks indices due to improvement in connectivity. Though its patterns may not exceed the level Greece may already have but the case of better connectivity and improved linkages must be taken into consideration. Beside general relationships it's interesting to understand the association between clustering coefficient and economic indicators. Unlike Greece it represents a negative link in between, depicting that it might be connecting more aggressively and removing the clustering barriers.

The basic relationship between GDP, PPP and centrality indices shows sign of positive correlation. There is a connection in-between. The conclusion for Tier 3 can be generalized for Tier 3 and Tier 2 countries as well. Improvements in connectivity and network position can increase or enhance the correlation patterns. The countries must strive to connect with all nodes to fully capitalize the opportunities for market efficiency and improvements on economic front.

**CONCLUSION**

The investment network is not complete graph. So connectivity pattern of different nodes varies accordingly. Some nodes have central position with higher connectivity levels compared to other nodes. The countries can be divided into different groups or tires based on the resultant centrality and analytical measures. The classification of countries into different groups explains the differences in connectivity patterns for nodes. We divided the nodes into three tires based on their closeness centrality. The connectivity is not uniformly distributed among all nodes as assumed in different earlier studies. Tier 1 countries have important position within network; with strong centrality indices we can find correlation patterns with their economic indicators. These correlations are not causal rather shows the existence of relationship. The general conclusion should include this trend. The more central a country is, there more relationships or correlations can be found. It can also have implications for countries with less connectivity to improve their network position and capture more economic benefits. The conclusion on representative country of tier 2 can be generalized for whole group due to economic similarities and prevailing circumstances. These nodes do not have strong connectivity pattern, so experience variations in their position within network. The links or correlation between network indices of Tier 2 countries and economic indicators is not high enough, due to limited connectivity and exposure. There basic relationship between GDP,

Purchasing Power Parity (PPP) and centrality indices shows sign of positive correlation. There is a connection in-between. The conclusion for Tier 3 can be generalized for Tier 3 countries. Improvements in connectivity and network position can increase or enhance the correlation patterns. The countries must strive to connect with all nodes to fully capitalize the opportunities for market efficiency and improvements on economic front. European Union is the case for other investment networks and individual countries to establish strong linkages to increase the connectivity patterns. The more connected nodes can have strong positive correlation between investment and economy.